\documentclass[12pt]{article}
\usepackage{putex}
\usepackage{latexsym}

\newcommand{\beq}{\begin{eqnarray}}
\newcommand{\eeq}{\end{eqnarray}}

\newcommand{\bi}{\bibitem}

\newcommand{\be}{\begin{equation}}
\newcommand{\ee}{\end{equation}}
\newcommand{\ben}{\begin{eqnarray}\displaystyle}
\newcommand{\een}{\end{eqnarray}}

\begin{document}

\preprint{hep-th/0409186 \\ PUPT-2134 \\ NSF-KITP-04-100}

\institution{PU}{Joseph Henry Laboratories,\cr Princeton University, Princeton, NJ 08544}
\institution{KITP}{Kavli Institute for Theoretical Physics,\cr University of California, Santa Barbara, CA  93106}

\title{Variations on the Warped Deformed Conifold}

\authors{Steven S.~Gubser,\worksat{\PU,}\footnote{e-mail: \tt 
ssgubser@Princeton.EDU} 
Christopher P.~Herzog,\worksat{\KITP,}\footnote{e-mail: \tt 
herzog@kitp.ucsb.EDU} 
and Igor R.~Klebanov\worksat{\PU,}\footnote{e-mail: \tt klebanov@Princeton.EDU}}

\abstract{
The warped deformed conifold background of type IIB theory is
dual to the cascading 
$SU(M(p+1))\times SU(Mp)$ gauge theory. 
We show that this background realizes the (super-)Goldstone mechanism
where the $U(1)$ baryon number symmetry is broken by
expectation values of baryonic operators. The resulting massless
pseudo-scalar and scalar glueballs are identified in the supergravity
spectrum. A D-string is then dual to a global string in the gauge theory.
Upon compactification, the Goldstone mechanism turns into the Higgs mechanism,
and the global strings turn into ANO strings.
}

\maketitle

\section{Introduction}

This talk, delivered at Strings '04 by one of us (I.R.K.),
is a condensed version of our paper \cite{GHK}.

One of the themes in recent string theory research concerns
extensions of the AdS/CFT correspondence 
\cite{Maldacena,Gubser,Witten:1998qj}
to confining gauge theories. One such background of type IIB string,
the warped deformed conifold, was constructed in \cite{KS}. 
It was argued to be dual to 4-dimensional ${\cal N}=1$ supersymmetric
$SU(M(p+1)) \times SU(Mp)$ gauge theory \cite{GK} whose flow exhibits an RG
cascade \cite{KS,KT}. In each cascade step the integer $p$ decreases by $1$
through the Seiberg duality \cite{Seiberg}.

In this talk we show that the warped deformed conifold
of \cite{KS} incorporates a supergravity dual of the supersymmetric Goldstone
mechanism, and identify a pseudo-scalar Goldstone boson and
its scalar superpartner.
An old puzzle guides our investigation: what is the
gauge theory interpretation of D1-branes in the deformed conifold
background \cite{KS}? 
The interpretation of the fundamental strings
placed in the IR region of the metric is clear:
they are dual to confining strings.
Like the fundamental strings, the D-strings fall to the bottom of
the throat, $\tau=0$, where they
remain tensionful; hence, they cannot be dual to `t Hooft loops which must
be screened \cite{KS}.
We propose instead that in the dual gauge theory they are 
solitonic strings
that create a monodromy of a massless pseudo-scalar Goldstone boson 
field.\footnote{ We are grateful to
E. Witten for emphasizing this possibility to us.}
For this explanation to make sense, the IR gauge theory must differ from the
pure glue ${\cal N}=1$ theory in that it contains a massless 
pseudo-scalar bound state (glueball).
The fact that this massless mode must couple directly to a D-string means that
it corresponds to a certain perturbation of the RR 2-form potential,
which turns out to mix with the RR 4-form potential.
We exhibit the necessary ansatz in section 3,
and indeed find a massless glueball. This mode should
be interpreted as the Goldstone boson of spontaneously
broken global $U(1)$ baryon number symmetry. 
Its presence supports the claim made in \cite{KS,Aharony} that
the cascading gauge theory is on the baryonic branch \cite{APS}, i.e. certain
baryonic operators acquire expectation values. 
The supersymmetric Goldstone mechanism gives rise also to a massless scalar
mode. In section 4 the supergravity dual of this mode is identified
as a massless glueball coming from a mixture of an NS-NS 2-form and
a metric deformation. The ansatz for such perturbations
was written down some time ago in \cite{PT}.

Besides being an interesting example of the gauge/gravity duality,
the warped deformed conifold background offers 
interesting possibilities for solving the hierarchy problem 
along the lines suggested in \cite{RS,Ver}.
If the background is embedded into a compact CY space with NS-NS and R-R
fluxes, then an exponential
hierarchy may be created between the UV compactification scale and
the IR scale at the bottom of the throat \cite{KS,GKP}. Models of this type received
an additional boost due to a possibility of fixing all moduli proposed in
\cite{KKLT}, and a subsequent exploration of cosmology in
\cite{KKLMMT}. Recently, 
a new role was proposed
for various $(p,q)$ strings placed in the 
IR region \cite{Copeland}. 
Besides being
the confining or solitonic strings from the point of view of the gauge theory, 
they
may be realizations of cosmic strings. The exponential warping of the
background lowers the tension significantly, and makes them 
plausible cosmic string candidates. 
In section 5 we discuss the Higgs mechanism that occurs
upon embedding the warped deformed conifold into a flux compactification,
and argue that a D-string placed at the bottom of
the throat is dual to an
Abrikosov-Nielsen-Olesen string in the 
gauge theory coupled to supergravity.
We conclude in section 6.

\section{Review of the Warped Deformed Conifold}

The conifold may be described by the
following equation in four complex variables,
\be \label{coni}
\sum_{a=1}^4 z_a^2 = 0
\ .
\ee
Since this equation is invariant under an overall real rescaling of
the coordinates, this space is a cone  and admits the metric \cite{Candelas}
\be
ds_6^2 = dr^2 + r^2 ds_{T^{1,1}}^2\ ,
\label{conimetric}
\ee
where
\begin{equation} \label{co}
ds_{T^{1,1}}^2=
{1\over 9} \bigg(d\psi +
\sum_{i=1}^2 \cos \theta_i d\phi_i\bigg)^2+
{1\over 6} \sum_{i=1}^2 \left(
d\theta_i^2 + {\rm sin}^2\theta_i d\phi_i^2
 \right)
\
\end{equation}
is the metric on $T^{1,1}$. Here $\psi$ is an angular coordinate
which  ranges from $0$ to $4\pi$,  while $(\theta_1,\phi_1)$
and $(\theta_2,\phi_2)$ parametrize two ${\bf S}^2$s in a standard way.
Therefore, this form of the metric shows that
$T^{1,1}$ is an ${\bf S}^1$ bundle over ${\bf S}^2 \times {\bf S}^2$.
Topologically, $T^{1,1}\sim {\bf S}^2 \times {\bf S}^3$.

Now placing $N$ D3-branes at the apex of the cone we find the metric
\begin{equation} \label{newgeom}
ds^2 = \left (1+{L^4\over r^4}\right )^{-1/2}
\left (- (dx^0)^2 +(dx^1)^2+ (dx^2)^2+ (dx^3)^2\right )  
\ee
\[
\qquad \qquad \qquad \qquad \qquad
+ \left (1+{L^4\over r^4}\right )^{1/2} (dr^2 + r^2 ds_{T^{1,1}}^2)\ ,
\]
whose near-horizon ($r\rightarrow 0$) limit is $AdS_5\times T^{1,1}$.
The same logic that leads us to the maximally supersymmetric
version of the AdS/CFT correspondence now shows that
the type IIB string theory on this space should
be dual to the infrared limit of the field theory on $N$ D3-branes
placed at the singularity of the conifold. Since Calabi-Yau spaces
with these D-branes 
preserve 1/4 of the original supersymmetries, we have
an ${\cal N}=1$
superconformal field theory.
This field theory was constructed
in \cite{KWit,MP}: it is $SU(N)\times SU(N)$ gauge theory
coupled to two chiral superfields, $A_i$, in the
$({\bf N}, \overline{\bf N})$
representation
and two chiral superfields, $B_j$, in the $(\overline{\bf N}, {\bf N})$
representation. 

The continuous symmetries of the gauge theory are $U(1)_R\times
U(1)_B\times SO(4)$ where the $SO(4)$ acts on the $A$'s and the $B$'s
as $SU(2)\times SU(2)$.  
The exactly marginal superpotential is
fixed uniquely by the symmetries 
up to overall normalization:
\be W\sim \epsilon^{ij}
\epsilon^{kl}\tr A_iB_kA_jB_l\ .
\ee
The $U(1)$ baryon number symmetry acts as
$A_k \to e^{i\alpha} A_k$, $B_j\to e^{-i\alpha} B_j$.
The massless gauge field in $AdS_5$ dual to the baryon number
current originates from the RR 4-form potential \cite{KW,Ceres}:
\be\label{gaugefield}
\delta C_4\sim \omega_3\wedge A \ .
\ee 

Also important for our discussion is the ${\bf Z}_2$ symmetry
generated by the interchange of $A_1,A_2$
with $B_1,B_2$ accompanied by charge conjugation, i.e. the interchange
of the fundamental and the antifundamental representations, in 
both $SU(N)$ gauge groups \cite{KWit,MP}. We will call this 
interchange symmetry the ${\cal I}$ symmetry.
The corresponding transformation in the IIB string theory on $AdS_5\times T^{1,1}$
is the interchange of
$(\theta_1,\phi_1)$ with $(\theta_2,\phi_2)$ (i.e., of
the two ${\bf S}^2$'s) accompanied by the $-I$ of the
$SL(2,{\bf Z})$ S-duality symmetry \cite{KWit,MP}. 
The action of the $-I$ of the $SL(2,{\bf Z})$ reverses the sign of
the NS-NS and R-R 2-form potentials,
$B_2$ and $C_2$.

The addition of $M$ fractional 3-branes (wrapped D5-branes) at the
singular point of the conifold changes the gauge group 
to $SU(N+M)\times SU(N)$ \cite{GK}.
The $M$ units of magnetic 3-form flux cause the conifold
to make a ``geometric transition'' to the deformed conifold
\begin{equation} \label{dconifold}
\sum_{a=1}^4 z_a^2 =
\epsilon^2\ ,
\end{equation}
in which the singularity of the conifold is removed
through the blowing-up of the  ${\bf S}^3$ of $T^{1,1}$.
Therefore, the dual of the cascading $SU(M(p+1))\times SU(Mp)$ gauge theory
is the warped deformed conifold \cite{KS}.
Below we collect some necessary formulae about this background (for
reviews see \cite{Herzog}).

The ten dimensional metric is
\be \label{10dmetric}
ds_{10}^2 = h(\tau)^{-1/2} \left (- (dx^0)^2 +(dx^1)^2+ (dx^2)^2+ (dx^3)^2\right )
+h(\tau)^{1/2} ds_6^2 \ ,
\ee
where
\be \label{conifoldmetric}
ds_6^2 = {\epsilon^{4/3} K(\tau) \over 2}
\Bigg[ {1 \over 3K^3} (d\tau^2 + (g_5)^2) 
 + \cosh^2 \left ({\tau\over 2}\right ) ((g^3)^2 + (g^4)^2)
\ee
\[
\qquad \qquad \qquad \qquad
+ \sinh^2 \left ({\tau\over 2}\right ) ((g^1)^2 + (g^2)^2) \Bigg] \ 
\]
is the usual Calabi-Yau metric on the deformed conifold.
The one forms are given in terms of angular coordinates as
\begin{eqnarray} \label{gbasis}
g^1 = {e^1-e^3\over\sqrt 2}\ ,\qquad
g^2 = {e^2-e^4\over\sqrt 2}\ , & \nonumber \\
g^3 = {e^1+e^3\over\sqrt 2}\ ,\qquad
g^4 = {e^2+ e^4\over\sqrt 2}\ , & \qquad
g^5 = e^5\ ,
\end{eqnarray}
where
\begin{eqnarray} \label{ebasis}
e^1\equiv - \sin\theta_1 d\phi_1 \ ,\qquad
e^2\equiv d\theta_1\ , 
%\nonumber \\
& \qquad
e^3\equiv \cos\psi\sin\theta_2 d\phi_2-\sin\psi d\theta_2\ , \nonumber \\
e^4\equiv \sin\psi\sin\theta_2 d\phi_2+\cos\psi d\theta_2\ , 
%\nonumber \\
& \qquad
e^5\equiv d\psi + \cos\theta_1 d\phi_1+ \cos\theta_2 d\phi_2 \ .
\end{eqnarray}
Note that 
\be \label{Keqn}
K(\tau)= { (\sinh (2\tau) - 2\tau)^{1/3}\over 2^{1/3} \sinh \tau}
\ .
\ee
The warp factor is
\be \label{intsol}
h(\tau) = (g_s M\alpha')^2 2^{2/3} \epsilon^{-8/3} I(\tau)\ ,
\ee
where
\be \label{Itau}
I(\tau) \equiv
\int_\tau^\infty d x {x\coth x-1\over \sinh^2 x} (\sinh (2x) - 2x)^{1/3}
\ .
\ee
Since $h(\tau)$ decreases monotonically from a finite value at $\tau=0$,
the tension of the fundamental string is minimized at $\tau=0$,
where it is found to be $1/(2\pi\alpha' \sqrt{h(0)})$. 
This means that this background is dual to
a confining gauge theory.

The NS-NS two form field is
\be \label{B2}
B_2 = {g_s M \alpha'\over 2} [f(\tau) g^1\wedge g^2
+  k(\tau) g^3\wedge g^4 ]\ ,
\ee
while the RR three form field strength is
\begin{eqnarray}
F_3 = {M\alpha'\over 2} \left \{g^5\wedge g^3\wedge g^4 + d [ F(\tau)
(g^1\wedge g^3 + g^2\wedge g^4)]\right \} 
\ .
\end{eqnarray}
The auxiliary functions in these forms are
\begin{eqnarray}
F(\tau) &=& {\sinh \tau -\tau\over 2\sinh\tau}\ ,
\nonumber \\
f(\tau) &=& {\tau\coth\tau - 1\over 2\sinh\tau}(\cosh\tau-1) \ ,
\nonumber \\
k(\tau) &=& {\tau\coth\tau - 1\over 2\sinh\tau}(\cosh\tau+1)
\ .
\end{eqnarray}

In the warped deformed conifold the $SO(4)$ and the ${\cal I}$
global symmetries
are preserved, but the $U(1)_R$ symmetry is broken
in the UV by the chiral anomaly down to ${\bf Z}_{2M}$ \cite{KOW}.
Further spontaneous breaking of this discrete
symmetry to ${\bf Z}_2$, which acts as $z_i \to - z_i$,
 does not lead to appearance of
a Goldstone mode. The $U(1)_B$ symmetry is not anomalous,
and its spontaneous breaking does produce 
a Goldstone mode, which we exhibit in section 3.

\section{The Goldstone mode}

To begin, consider a D1-brane extended in two of the four dimensions
in ${\bf R}^{3,1}$.
Because the D1-brane carries electric charge under the 
R-R three-form field strength $F_3$, it is natural to think that a 
pseudo-scalar $a$ in four dimensions, 
defined so that $*_4 da = \delta F_3$,\footnote{
The 4-dimensional Hodge dual $*_4$ is calculated with the Minkowski metric,
${\rm vol}_4= dx^0 \wedge dx^1\wedge dx^2\wedge dx^3$.} 
experiences monodromy as one loops around the D1-brane world-volume.  

The perturbation ansatz we therefore adopt is
\begin{eqnarray} \label{NewPerturbation}
  \delta F_3 &=& *_4 da + 
   f_2(\tau) da \wedge dg^5 + f_2' da \wedge d\tau \wedge g^5\ ,  \nonumber \\
  \delta F_5 &=& (1 + *) \delta F_3 \wedge B_2=
(*_4 da
- {\epsilon^{4/3}\over 6 K^2(\tau)} h(\tau) da\wedge d\tau\wedge g^5) \wedge B_2 \,,
 \end{eqnarray}
where $f_2'= d f_2/d\tau$, $h(\tau)$ is
given by (\ref{intsol}), and $K(\tau)$ by (\ref{Keqn}).  
The variations of all other fields, including the metric and the dilaton, vanish.
The last two terms in 
$\delta F_3$ sum to the exact form $-d (f_2 da \wedge g^5)$. 
As shown in \cite{GHK}, all linearized SUGRA equations are satisfied if
$a(x^0,x^1,x^2,x^3)$ is a harmonic function, i.e. $d*_4 da = 0$,
and $f_2(\tau)$ satisfies
\begin{equation}\label{threef}
-{d\over d\tau} [K^4 \sinh^2 \tau f_2'] + {8\over 9 K^2} f_2 =
{ (g_s M \alpha')^2\over 3 \epsilon^{4/3} } (\tau\coth \tau -1)\left (\coth\tau -
{\tau\over \sinh^2\tau}\right ) \,.
\end{equation}
The normalizable solution of (\ref{threef})
that is regular both for small
and for large $\tau$ is
 \be \label{FoundFTwo}
  f_2 (\tau) 
=
   -{2 c \over K^2 \sinh^2 \tau}\int_0^\tau dx \, h(x) \sinh^2 x \,,
\ee
where $c \sim \epsilon^{4/3}$.
We find that
$f_2\sim \tau$ for small $\tau$, and $f_2\sim \tau e^{-2\tau/3}$ for large $\tau$.

The zero-mass glueball we are finding is due to the spontaneously
broken global $U(1)$ baryon number symmetry \cite{Aharony}.
 The form of the
$\delta F_5$ in (\ref{NewPerturbation}) makes the connection between
our zero-mode and $U(1)_B$ evident.
Asymptotically, at large $\tau$, 
there is a component $\sim \omega_3\wedge da\wedge d\tau$ in $\delta F_5$.
Thus from (\ref{gaugefield}), we have 
$A\sim da$. For the 4-d effective Lagrangian,
there should be a coupling between the baryon number current $J^\mu$
and $a$:
\begin{equation} \label{Jda}
{1\over f_a}\int d^4 x J^\mu \partial_\mu a 
=-{1\over f_a}\int d^4 x \, a(x) (\partial_\mu J^\mu) \ ,
\end{equation}
where the pseudo-scalar $a$ enters as the parameter of the baryon number
transformation. 
 It is important that this transformation
does not leave the vacuum invariant!

As discussed in \cite{KS,Aharony}
the field theory is on the baryonic branch: 
``the last step'' of the cascade takes place through
giving expectation values to baryonic operators in the
$SU(2M)\times SU(M)$ gauge theory
coupled to bifundamental fields
$A_i, B_j$, $i,j=1,2$. 
In addition to mesonic operators
$(N_{ij})^\alpha_\beta \sim (A_i B_j)^{\alpha}_\beta$, the gauge theory has 
baryonic operators invariant under the $SU(2M)\times SU(M)$
gauge symmetry:
\begin{eqnarray} \label{baryonops}
{\cal B}& \sim  & \epsilon_{\alpha_1\alpha_2\ldots \alpha_{2M}}
(A_1)^{\alpha_1}_1 (A_1)^{\alpha_2}_{2}\ldots
(A_1)^{\alpha_M}_{M} 
(A_2)^{\alpha_{M+1} }_{1} (A_2)^{\alpha_{M+2} }_{2}\ldots
(A_2)^{\alpha_{2M} }_{M} \ ,\cr
\bar {\cal B}& \sim & \epsilon^{\alpha_1\alpha_2\ldots
\alpha_{2M}}
(B_1)^{1}_{\alpha_1} (B_1)^{2}_{\alpha_2}\ldots
(B_1)^{M}_{\alpha_M} 
(B_2)^{1}_{\alpha_{M+1} } 
(B_2)^{2}_{\alpha_{M+2} }\ldots 
(B_2)^{M}_{\alpha_{2M} }\ . 
\end{eqnarray}
The baryonic operators are invariant under
the $SU(2)\times SU(2)$ global symmetry rotating $A_i,B_j$.
These operators contribute an additional
term to the usual mesonic superpotential:
\begin{equation}
W = \lambda (N_{ij})^\alpha_\beta
(N_{k\ell})_\alpha^\beta\epsilon^{ik}\epsilon^{j\ell}
+ X(\det [(N_{ij})^\alpha_\beta]
-{\cal B}\bar{\cal B} - \Lambda_{2M}^{4M}) \ ,
\end{equation}
where $X$ can be understood as a Lagrange multiplier.

The supersymmetry-preserving vacua include the baryonic branch:
\be \label{superW}
X = 0 \ ; \ N = 0 \ ;\quad 
 {\cal B} \bar{\cal B} = -\Lambda_{2M}^{4M} \ ,
\ee
where the $SO(4)$ global symmetry 
rotating $A_i,B_j$
is unbroken.  Since the supergravity background of \cite{KS}
also has this symmetry, it is natural to identify
the dual of this background with the baryonic branch of
the cascading theory. 
The expectation values of the baryonic operators
spontaneously break the $U(1)$ baryon number symmetry
$A_k \to e^{i\alpha} A_k$, $B_j\to e^{-i\alpha} B_j$.
The deformed conifold as described in \cite{KS} 
corresponds to a vacuum where 
$|{\cal B}| = |\bar{\cal B}|=\Lambda_{2M}^{2M}$, which is
invariant under the exchange of the
$A$'s with the $B$'s accompanied by charge conjugation
in both gauge groups. As noted in \cite{Aharony}, 
the baryonic branch has complex dimension $1$, 
and it can be parametrized by $\xi$ where
 \be \label{xiDef}
  {\cal B} = i\xi \Lambda_{2M}^{2M} \,,\qquad
  \bar{\cal B} = {i \over \xi} \Lambda_{2M}^{2M} \,.
 \ee
The pseudo-scalar Goldstone mode must correspond to 
changing $\xi$ by a phase, 
since this is precisely what a $U(1)_B$ symmetry transformation does.  
As usual, the gradient of
the pseudo-scalar Goldstone mode 
$f_a\partial_\mu a$ is created from the vacuum by the action of 
the axial baryon number current, $J_\mu$ (we expect that the scale
of the dimensionful `decay constant' $f_a$ is determined by the
baryon expectation values).

Thus, the breaking of the $U(1)$ baryon number symmetry
necessitates the presence of a massless 
pseudo-scalar glueball, which we have found.
By supersymmetry, this field falls into a massless ${\mathcal N}=1$
chiral multiplet.  
Hence, there will also be a massless 
scalar mode 
and corresponding Weyl fermion. The scalar must
correspond to changing $\xi$ by a positive real factor.

\section{The Scalar Zero-Mode}
\label{Scalar}

The presence of the pseudo-scalar zero mode found in section 3, 
and the ${\cal N}=1$ supersymmetry, require the existence of a
scalar zero-mode.  In this section we argue that this zero-mode comes from
a metric perturbation that mixes with the NS-NS 2-form
potential.

The warped deformed conifold of \cite{KS} preserves the
 ${\bf Z}_2$ interchange symmetry which we called the ${\cal I}$
symmetry in section 2: see below (\ref{gaugefield}).
However, the pseudo-scalar mode we found breaks this 
symmetry: 
from the form of the 
perturbations (\ref{NewPerturbation}) we see that 
$\delta F_3$ is even under the interchange of 
$(\theta_1,\phi_1)$ with $(\theta_2,\phi_2)$,  while $F_3$
is odd; $\delta F_5$ is odd while $F_5$ is even.
Similarly, the scalar mode must also break the ${\cal I}$ symmetry
because in the field theory it breaks the symmetry between
expectation values of $|{\cal B}|$ and of $|\bar {\cal B}|$.
We expect that turning on the zero-momentum scalar
modifies the geometry because 
the scalar changes the absolute value of  
$|{\cal B}|$ and $|\bar {\cal B}|$ while the pseudo-scalar affects only the phase.
The necessary perturbation that preserves the $SO(4)$ but breaks the
${\cal I}$ symmetry
is a mixture of the NS-NS 2-form
and the metric:
\be \label{NSPerturbation}
  \delta B_2 =  \chi(\tau) dg^5\ , \; \; \; \; \;
\delta G_{13}  =  \delta G_{24} = 
m(\tau)\ , 
\ee
where, for example, $\delta G_{13} = m(\tau)$ means to add 
$2m(\tau)\, g^{(1} g^
{3)}$ to $ds_{10}^2$.
To see that these components of the metric break the ${\cal I}$
symmetry, we note that
\begin{equation}
(e^1)^2 + (e^2)^2 - (e^3)^2 - (e^4)^2 =
g^1 g^3 + g^3 g^1+ g^2 g^4+ g^4 g^2\ . 
\end{equation}
We  find it convenient to define
%\begin{equation} \label{ztaudef}
$m(\tau)= h^{1/2} K\sinh (\tau) \, z (\tau)$.
%=
%2^{-1/3} [\sinh(2\tau) - 2\tau]^{1/3} h^{1/2} z (\tau)
%\ .
%\end{equation}
%Then (\ref{raiseform}) becomes
%\eqn{raiseformnew}{ 
%\chi' =
%g_s M z(\tau) [f' \coth(\tau/2)  + k' \tanh(\tau/2) ]
%={1\over 2} g_s M z(\tau) {\sinh(2\tau) - 2\tau\over \sinh^2 \tau}
%\ .
%} 
%This is equivalent to (5.21) of \cite{PT}.

%Equations~\eno{starH} and~\eno{Hupper} also imply that $\delta (*_6 H_3)=0$,
%where $*_6$ is calculated with the 6-d CY metric.
%And with \raiseform\ in hand, one may straightforwardly show that 
%$\delta(*_6 F_3) = -{1 \over g_s} \delta H_3$, where the left hand side 
%is non-vanishing even though $\delta F_3 = 0$ because the definition of 
%Hodge duals changes when the metric is varied.  
%The upshot is that the variation we are considering preserves the 
%self-duality condition of the 3-forms, $*_6 H_3 =-g_s F_3$.

In \cite{GHK} it was shown that all the linearized SUGRA equations are
satisfied provided that
\be \label{zdiffeq}
\frac{\left( \left( K \sinh(\tau) \right)^2 z' \right)' }
{(K \sinh(\tau))^2 } = \left(
\frac{2}{\sinh(\tau)^2} +\frac{8}{9} \frac{1}{K^6 \sinh(\tau)^2}
- \frac{4}{3} \frac{\cosh(\tau)}{K^3 \sinh(\tau)^2}
\right) z \ 
\ee
and
\be
\chi' =
%g_s M z(\tau) [f' \coth(\tau/2)  + k' \tanh(\tau/2) ]
{1\over 2} g_s M z(\tau) {\sinh(2\tau) - 2\tau\over \sinh^2 \tau} \ .
\ee
The solution of (\ref{zdiffeq}) for the zero-mode
is remarkably simple:
\be \label{zresult}
z(\tau)=s {
(\tau\coth (\tau)-1) \over
[\sinh(2\tau)-2\tau]^{1/3} }
\ ,
\ee
with 
$s$ a constant.
Like the pseudo-scalar perturbation, the large $\tau$ asymptotic is again 
$z\sim \tau e^{-2\tau/3}$. 
We note that the metric perturbation also has the simple form
$\delta G_{13}\sim h^{1/2} [\tau\coth (\tau)-1]
$.
Note that the perturbed metric
$d\tilde s^6_2$ differs from the metric of the deformed
conifold, eq.~(\ref{conifoldmetric}), by
\be
\sim (\tau\coth\tau -1) (g^1 g^3 + g^3 g^1 + g^2 g^4 + g^4 g^2)
\ ,
\ee
which grows as $\ln r$ in the asymptotic radial variable $r$.

The existence of the
scalar zero-mode makes it likely
that there is a one-parameter family of supersymmetric solutions
which break the ${\cal I}$ symmetry but preserve the $SO(4)$
(an ansatz with these properties was found in \cite{PT}, and
its linearization agrees with (\ref{NSPerturbation})).
We will call these conjectured backgrounds
{\bf resolved warped deformed conifolds}.
We add the word {\bf resolved} because 
both the resolution of the conifold, which is a Kaehler deformation,
and these resolved warped deformed conifolds
break the ${\cal I}$ symmetry.
As we explained in section 3, in the dual gauge theory
turning on the ${\cal I}$ breaking corresponds to
the transformation
${\cal B} \to (1+s) {\cal B}$,
$\bar{\cal B} \to (1+s)^{-1} \bar{\cal B}$ on the baryonic branch.
Therefore, $s$ is dual to the ${\cal I}$ breaking parameter of the 
resolved warped deformed conifold.

One might ask whether the resolved warped deformed conifolds
are still of the form 
$h^{-1/2} dx_{||}^2 + h^{1/2} d\tilde s_6^2$
where $d\tilde s_6^2$ is Ricci flat.  
At linear 
order in our perturbation, our conifold
metric $d\tilde s_6^2$ is indeed Ricci flat: the first order corrections
vanish if (\ref{zdiffeq}) is satisfied.
We also showed 
%(see below \raiseformnew) 
\cite{GHK}
that the complex 3-form field
strength $G_3= F_3- {i\over g_s} H_3$ remains
imaginary self-dual at linear order,
i.e.~$*_6 G_3 =i G_3$.
It will be interesting to see if these properties continue to hold for the
exact solution.

\section{Compactification and Higgs Mechanism}

As we argued above, the non-compact warped deformed conifold
exhibits a supergravity
dual of the Goldstone mechanism. It was crucial for our arguments
that the $U(1)_B$ symmetry is not gauged in the field theory, and   
the appearance of the Goldstone boson in the supergravity dual confirms that
the symmetry is global.

If the warped deformed conifold is embedded into a flux compactification
of type IIB string on a 6-dimensional CY manifold, then we expect
the global $U(1)_B$ symmetry to become gauged, because the square
of the gauge coupling becomes finite.
In the compact case we may write $\delta C_4 \sim
\omega_3 \wedge A$, where $\omega_3$ is harmonic in the full compact
case and $A$ is the 4-d gauge field. If we ignore subtleties with the self-duality
of the 5-form field strength, then the kinetic terms for it is
\be
{1\over 2 g_s^2} \int d^{10} x \sqrt{-g} F_5^2\ .
\ee
Substituting $F_5=F_2\wedge \omega_3$ and reducing to 4 dimensions, we
find the $U(1)$ kinetic term
\be
{1\over 2 g^2} \int d^4 x F_2^2\ ,
\ee
where
\be
{1\over g^2} \sim {1\over g_s^2} \tau_m\ ,
\ee
where we assumed that the effect of compactification is to introduce a cut-off
at $\tau_m\gg 1$.

The finiteness of the gauge coupling in the compact case means that
the Goldstone mechanism should turn into a Higgs mechanism.
The Goldstone boson $a$ enters as a gauge parameter of
$A$ and gets absorbed by the $U(1)$ gauge field to make a
massive vector field. As usual in the supersymmetric Higgs mechanism,
the scalar acquires the same mass which originates from the D-term
potential.
In ${\cal N}=1$ notation, 
gauge invariance means we have to introduce factors of $e^{\pm g V}$
into the D-terms for ${\cal B}$ and $\overline {\cal B}$:\footnote{
${\cal B}$ and $\overline {\cal B}$ have charge of order $M$, a charge which
we have for simplicity neglected to include in the coupling to $V$.
} 
\be
{\cal B}^* e^{g V} {\cal B} + \overline{\cal B}^* e^{-g V} \overline{\cal B} \ .
\ee
Expanding these D-terms to second order in $g$, we find
\be
g^2 (|{\cal B}|^2 + |\overline {\cal B}|^2) V^2  \ .
\ee
As a result, we find an
${\cal N}=1$ massive vector supermultiplet containing
a massive vector, a scalar (the Higgs boson), and their fermion   
superpartners.

% SSG
In the preceding, we ignored the linear term in $g$:
\be
g (|{\cal B}|^2 - |\overline {\cal B}|^2 + \zeta) V \ ,
\ee
where we have included a Fayet-Iliopoulos parameter $\zeta$.
Depending on details of the compactification,
it may be that $\zeta$ is nonzero.
Then 
the potential for the scalar has its minimum for
$|\xi| \neq 1$, and the baryon VEVs in (\ref{xiDef})
are unequal in magnitude. In other words, the 
Fayet-Iliopoulos parameter breaks the ${\cal I}$ symmetry.
Thus, further study of the one-parameter family of the
supersymmetric backgrounds dual to the baryonic branch is
of interest to the understanding of flux compactifications.  

While in the non-compact case D-strings are global strings,
in the compact case they should be interpreted as
Abrikosov-Nielsen-Olesen vortices of an Abelian-Higgs
model, where the charged chiral superfields
breaking the gauge symmetry are the baryon operators ${\cal B}$  
and $\bar {\cal B}$.\footnote{
Representation of D-strings by ANO vortices
in low-energy supergravity was recently advocated in 
a different context \cite{Dvali}
(see also earlier work by \cite{Edelstein}).}
Since there is
a finite number $K$ of NS-NS
flux units through a cycle dual to the 3-sphere \cite{GKP},
 the D-string charge
takes values in ${\bf Z}_K$. Indeed, $K$ D-strings can break
on a wrapped D3-brane \cite{Copeland}. Correspondingly, we
do not expect the ANO vortex duals to be BPS saturated.

\section{Discussion}

Our work sheds new light on the physics of the cascading 
$SU(M(p+1))\times SU(Mp)$ gauge theory, whose supergravity dual is the warped
deformed conifold \cite{KS}. In the infrared the theory is not 
in the same universality class as the pure glue ${\cal N}=1$
supersymmetric $SU(M)$ theory: the cascading theory contains
massless glueballs, as well as solitonic strings dual to
the D-strings placed at $\tau=0$ in the supergravity
background. 

As suggested in \cite{KS,Aharony} and reviewed in section 3 above, the infrared field theory 
is better thought of as $SU(2M)\times SU(M)$ on the baryonic branch, i.e.
with baryon operators (\ref{baryonops}) having expectation values.
Since the global baryon number symmetry, $U(1)_B$, 
is broken by these expectation values, 
the spectrum must contain a Goldstone bosons which 
we find explicitly. 
We also construct at linear order a Lorentz-invariant 
deformation of the background which we argue is a zero-momentum 
state of the scalar superpartner of the Goldstone mode.  
Our calculations confirm the validity 
of the baryonic branch interpretation of the gauge theory.
This also resolves a puzzle about the dual of the D-strings
at $\tau=0$: they are the solitonic strings that couple to these
massless glueballs. We further argue that, upon embedding
this theory in a warped compactification, the global
$U(1)_B$ symmetry becomes gauged; then the gauge symmetry
is broken by the baryon expectation values through a
supersymmetric version of the
Higgs mechanism. Thus, in a flux compactification,
we expect the D-string to be dual to an Abrikosov-Nielsen-Olesen
vortex.  

In \cite{KS} it was argued that there is a limit, $g_s M\to 0$,
\footnote{
No string theoretic description of this limit is yet available, because it
is the opposite of the limit of large $g_s M$ where the supergravity
description is valid. }
where the physics of the cascading gauge theory should approach that of the
pure glue ${\cal N}=1$ supersymmetric $SU(M)$ gauge theory.
How can this statement be consistent with the presence
of the Goldstone bosons? We believe that it can. 
Returning to the $SU(2M)\times SU(M)$ gauge theory discussed
in section 4, we expect that
in the limit $g_s M\to 0$ the scale $\Lambda_{2M}$ of the
$SU(2M)$, i.e. that of the baryon condensates,
is much higher than the scale $\Lambda_M$ of the $SU(M)$. 
Hence, the decay constant $f_a$ should be much greater
than the confinement scale $\Lambda_M$. Since the
Goldstone boson interactions at the confinement
scale are suppressed by powers of $\Lambda_M/f_a$, 
they appear to decouple from the massive glueballs containing
the physics of the pure glue supersymmetric $SU(M)$ gauge theory.
Obviously, this heuristic argument needs to be subjected
to various checks.

Our work opens new directions for future research. 
Turning on finite scalar perturbations is expected to give rise to
a new class of Lorentz invariant supersymmetric backgrounds,
{\bf the resolved warped deformed conifolds}, which preserve the $SO(4)$
global symmetry but break the discrete ${\cal I}$ symmetry of the
warped deformed conifold.
The ansatz for such backgrounds was proposed in
\cite{PT}. 
We have argued that these conjectured backgrounds are dual to the
cascading gauge theory on the baryonic branch. 
It would be desirable to find them explicitly,
and to confirm their supersymmetry.

A more explicit construction of the solitonic string 
in the gauge theory is desirable. 
It is also interesting to explore the 
consequences of our results for cosmological modeling.

%%%%%%%%%%%%%%%%%%%%%%%%%%%%%%%%%%%%%%%%%%%%%%%
\section*{\large Acknowledgments}
%%%%%%%%%%%%%%%%%%%%%%%%%%%%%%%%%%%%%%%%%%%%%%%%%%%%
We are grateful to D. Berenstein, J. Maldacena, 
J. Polchinski, A. Polyakov, R. Roiban, N. Seiberg, and especially 
O. Aharony, M. Strassler, and E. Witten for useful
discussions. IRK thanks the organizers of the conference
Strings '04 for their hospitality.
The work of SSG was supported in part by the Department of 
Energy under Grant No.\ DE-FG02-91ER40671, and by the Sloan Foundation.
The work of CPH was supported in part by
the National Science Foundation Grant No. PHY99-07949.
  The work of IRK was supported in part by
the National Science Foundation Grants No.
PHY-0243680 and PHY-0140311.
Any opinions, findings, and conclusions or recommendations expressed in
this material are those of the authors and do not necessarily reflect
the views of the National Science Foundation.

\end{document}